\documentclass[11pt]{article} 
\usepackage{hyperref} 
\pdfoutput=1

\begin{document} 
 
\title{Scrambled and Unscrambled Turbulence} 
 
\author{P. Ramaprabhu, V. Karkhanis and A.G.W. Lawrie \\ 
\\\vspace{6pt} University of  North  Carolina at Charlotte, NC 28078,USA \\ University of Bristol, United Kingdom} 
 
\maketitle 

\begin{abstract} 
The linked fluid dynamics videos depict Rayleigh-Taylor turbulence when driven by a complex acceleration profile involving two stages of acceleration interspersed with a stage of stabilizing deceleration. Rayleigh-Taylor (RT) instability occurs at the interface separating two fluids of different densities, when the lighter fluid is accelerated in to the heavier fluid. The turbulent mixing arising from the development of the miscible RT instability is of key importance in the design of Inertial Confinement Fusion capsules, and to the understanding of astrophysical events, such as Type Ia supernovae. By driving this flow with an ‘accel-decel-accel’ profile, we have investigated how structures in RT turbulence are affected by a sudden change in the direction of the acceleration first from destabilizing acceleration to deceleration, and followed by a restoration of the unstable acceleration. By studying turbulence under such highly non-equilibrium conditions, we hope to develop an understanding of the response and recovery of self-similar turbulence to sudden changes in the driving acceleration. 
\end{abstract} 
 
 
\section{Introduction} 
 
Detailed simulations of RT turbulence under a constant acceleration and the time-dependent accel-decel-accel profile were performed by MOBILE, a parallelized, three-dimensional, variable-density, finite volume incompressible Navier-Stokes solver [1]. The simulations were performed at a resolution of 512 x 512 x 2048 zones, with a computational domain elongated along the direction of the applied acceleration. The simulations were perturbed with an initial spectrum consisting of an annular shell of energetic modes in wavenumber space, so that the large scales that would drive the flow at late time are generated by coupling of the initial modes that were specified in the initial conditions. Periodic boundary conditions were enforced in the homogeneous directions, while the vertical surfaces were treated as outflow boundaries.\\
The first clip of RT turbulence shows the time evolution of isosurfaces of density from a 3D simulation in which the acceleration was held constant. As a result, the flow evolves to yield self-similarity, anisotropy in the direction of the applied acceleration, while large structures that dominate the flow at late time are aggregated through coupling of smaller structures constituting an inverse cascade process. Density contours realized at a horizontal midplane clearly highlight the inverse cascade process, as the flow evolves from high-wavenumber initial perturbations to low-wavenumber plumes evident at late times. Naturally, this process is accompanied by significant mixing between the two fluids.\\
When the flow is driven by the accel-decel-accel profile, the evolution is significantly more complex as evident first in the 3D density isosurface clip, as well as the density contours realized at the horizontal midplane. At the onset of the deceleration, previously growing coherent structures respond by reversing direction. During this stage, the structures are driven by inertia, and collide with each other breaking up in to smaller fragments. This ‘shredding’ (or scrambling) of large scale structures in to smaller fragments is accompanied by a sudden increase in mixing evident by the homogenization of density contours on the horizontal plane. Correspondingly, the 3D density iso-surfaces display a collapse of the mixing layer (pancaking) due to the reversal in direction of the coherent structures followed by the collision process. Interestingly, when the destabilizing acceleration is restored, the instability recovers (unscrambling) as the late-stage flow-features resemble the flow field from the constant acceleration simulation. This includes a restoration of anisotropy, self-similarity and the inverse cascade that drives the large-scale development. All of these features are evident in the evolution of power spectra of density fluctuations computed on the horizontal plane: the initial growth showing a significant inertial range, collapse in fluctuation energy during deceleration and shredding, and restoration of spectra with -5/3 profile at late times. In [2], we have investigated and report on key turbulent statistics during the stable deceleration and unstable reacceleration phases. \\
The video described here can be seen at the following URLs: \\
\href{URL of video}{Video 1 – Low resolution}\\
\href{URL of video}{Video 2 – High resolution}\\
This video has been submitted to the Gallery of Fluid Motion 2013 which is an annual showcase of fluid dynamics videos. \\
This work was supported in part by the (U.S.) Department of Energy (DOE) under Contract No. DE-AC52-06NA2-5396. AGWL would like to acknowledge the support of Ecole Centrale de Lyon. The simulations reported here were performed on Kraken at the National Institute for Computational Sciences, an advanced computing resource supported by the National Science Foundation.

References:\\
1. A.G.W.Lawrie and S. B. Dalziel, “Turbulent diffusion in tall tubes I. Models for Rayleigh-Taylor instability" Phys. Fluids 23, 085109, (2011). 
2. P. Ramaprabhu, V. Karkhanis and A.G.W. Lawrie, “The Rayleigh-Taylor Instability driven by an accel-decel-accel profile”, Submitted to Phys. Fluids (2013).

\end{document}